\documentclass[sn-mathphys]{sn-jnl}

\jyear{2021}%

\theoremstyle{thmstyleone}%
\theoremstyle{thmstyletwo}%

\theoremstyle{thmstylethree}%
\usepackage{soul}

\usepackage[normalem]{ulem}
\usepackage{amsfonts} 

\usepackage{xcolor}
\begin{document}

\title[Self-supervised learning for paranasal anomaly classification]{Self-supervised learning for classifying paranasal anomalies in the maxillary sinus}


\author[1,2]{Debayan Bhattacharya}
\author[1]{Finn Behrendt}
\author[2]{Benjamin Tobias Becker}
\author[1]{Lennart Maack}
\author[3]{Dirk Beyersdorff}
\author[4]{Elina Petersen}
\author[5]{Marvin Petersen}
\author[5]{Bastian Cheng}
\author[2]{Dennis Eggert}
\author[2]{Christian Betz}
\author[2]{Anna Sophie Hoffmann*}
\author[1]{Alexander Schlaefer*}

\affil[1]{Institute of Medicial Technology and Intelligent Systems, Technische Universitaet Hamburg, Germany}
\affil[2]{ 
Department of Otorhinolaryngology, Head and Neck Surgery and Oncology}
\affil[3]{ 
Clinic and Polyclinic for Diagnostic and Interventional Radiology and Nuclear Medicine}
\affil[4]{ Population Health Research Department, University Heart and Vascular Center}
\affil[5]{ Clinic and Polyclinic for Neurology}
\affil{University Medical Center Hamburg-Eppendorf, Hamburg, Germany}



 \abstract{
\footnotetext{*Equal contribution}
\footnotetext{For correspondence send email to: debayan.bhattacharya@tuhh.de}

\textbf{Purpose:} Paranasal anomalies, frequently identified in routine radiological screenings, exhibit diverse morphological characteristics. Due to the diversity of anomalies, supervised learning methods require large labelled dataset exhibiting diverse anomaly morphology. Self-supervised learning (SSL) can be used to learn representations from unlabelled data. However, there are no SSL methods designed for the downstream task of classifying paranasal anomalies in the maxillary sinus (MS).

\textbf{Methods:} Our approach uses a 3D Convolutional Autoencoder (CAE) trained in an unsupervised anomaly detection (UAD) framework. Initially, we train the 3D CAE to reduce reconstruction errors when reconstructing normal maxillary sinus (MS) image. Then, this CAE is applied to an unlabelled dataset to generate coarse anomaly locations by creating residual MS images. Following this, a 3D Convolutional Neural Network (CNN) reconstructs these residual images, which forms our SSL task. Lastly, we fine-tune the encoder part of the 3D CNN on a labelled dataset of normal and anomalous MS images.

\textbf{Results:} The proposed SSL technique exhibits superior performance compared to existing generic self-supervised methods, especially in scenarios with limited annotated data. When trained on just 10\% of the annotated dataset, our method achieves an Area Under the Precision-Recall Curve (AUPRC) of 0.79 for the downstream classification task. This performance surpasses other methods, with BYOL attaining an AUPRC of 0.75, SimSiam at 0.74, SimCLR at 0.73 and Masked Autoencoding using SparK at 0.75.

\textbf{Conclusion:} A self-supervised learning approach that inherently focuses on localizing paranasal anomalies proves to be advantageous, particularly when the subsequent task involves differentiating normal from anomalous maxillary sinuses. Access our code at \href{https://github.com/mtec-tuhh/self-supervised-paranasal-anomaly}{https://github.com/mtec-tuhh/self-supervised-paranasal-anomaly}

}




\maketitle

\section{Introduction}

The paranasal sinuses, air-filled spaces within the craniofacial complex, vary significantly and include the maxillary, frontal, sphenoid, and ethmoid sinuses \cite{9910059601802121}. Common pathologies like retention cysts, polyps, and mucosal thickening are identifiable through radiological screenings \cite{Bal2014-uw, Varshney2015-qi, Van_Dis1994-xz}. However, their diagnosis is challenging due to their incidental nature and the variability in sinus appearance \cite{Hansen.2014}. Research underscores their prevalence and the importance of accurate diagnosis in patient care \cite{Tarp.2000}. 3D imaging from computed tomography (CT) and magnetic resonance images (MRI) is vital for precise diagnosis, and misdiagnosis can lead to patient distress and increased healthcare costs \cite{Brierley.2017, Gutmann.2013}. The anatomical variability of the sinuses \cite{Papadopoulou2021-ch} necessitates careful application of deep learning for reliable diagnoses .

Convolutional Neural Networks (CNN) are recognized for diagnosing paranasal pathologies, evidenced in sinusitis classification \cite{Jeon.2021,Kim.2019}, differentiating inverted papilloma from carcinomas \cite{Liu.2022}, and detecting MS fungal ball and chronic rhinosinusitis in CT scans \cite{10.1371/journal.pone.0263125}. Prior studies have explored contrastive learning and cross-entropy loss for MS anomaly classification \cite{10.1007/978-3-031-16437-8_41}, and MS extraction techniques from MRI \cite{bhattacharya2023multiple}. However, all of the aforementioned methods use supervised learning. Given the difficulty in obtaining well-labelled datasets in clinical settings \cite{10.1145/3439950}, and the relative ease of acquiring unlabelled data, self-supervised learning (SSL), which learns representations from unlabelled data to improve the downstream task, has not yet been explored for paranasal anomaly classification. SSL efficiently utilizes unlabelled data through tasks like non-linear compression \cite{9412239,Xie2023}, denoising \cite{10.5555/1756006.1953039}, feature alignment from augmented images \cite{NEURIPS2020_f3ada80d,9578004,Huang2023} and \textcolor{black}{inpainting masked regions of images \cite{tian2023designing}}. However, these methods are \textcolor{black}{designed to improve the performance of models exposed to 2D natural images. Hence, they} lack a specific focus on enhancing MS anomaly classification from 3D MRI. \textcolor{black}{Our aim is to design an SSL task that enables the models trained on it to achieve maximum data efficiency in classifying paranasal anomalies.} We hypothesize anomaly segmentation within MS could be a good SSL task. Without ground truth segmentation masks, we use a UAD framework, applied in brain \cite{Baur2021-uz, 9761443} and paranasal anomaly detection \cite{https://doi.org/10.48550/arxiv.2211.01371}, to localize MS anomalies. A 3D Convolutional Autoencoder (CAE) trained on a labelled \textit{normal} dataset is used to reconstruct MS volumes and localize anomalies in an unlabelled dataset by failing to reconstruct anomalies leading to reconstruction errors. These errors, serving as pseudo segmentation masks are used in the SSL task to localize anomalies. We investigate if a 3D CNN, predicting these errors as SSL task, can improve feature discrimination between anomalous and normal MS in our labelled dataset. Our SSL task leverages available normal MS data, essential for supervised downstream task training.

Overall, our main contributions can be summed up as follows: 
\begin{itemize}
    \item  We present a self-supervised method that improves the downstream classification of normal vs anomlous MS. Our self-supervision task explicitly learns to coarsely localize anomalies by reconstructing the residual volumes generated through the UAD-trained autoencoder. This distinguishes our approach from the compared methods, where anomaly localization is not a primary focus for the self-supervision task. 
    
    \item Our self-supervised method effectively utilizes labelled healthy MS data reserved for downstream tasks. Hence, we explore how varying the CAE training set impacts downstream classification performance.

    \item We investigate post-processing strategies and loss function used in the self-supervision task for learning better transferable features for the downstream task. 
\end{itemize}

\section{Methods}

\begin{figure}[htbp]
  \centering
  \includegraphics[width=\columnwidth]{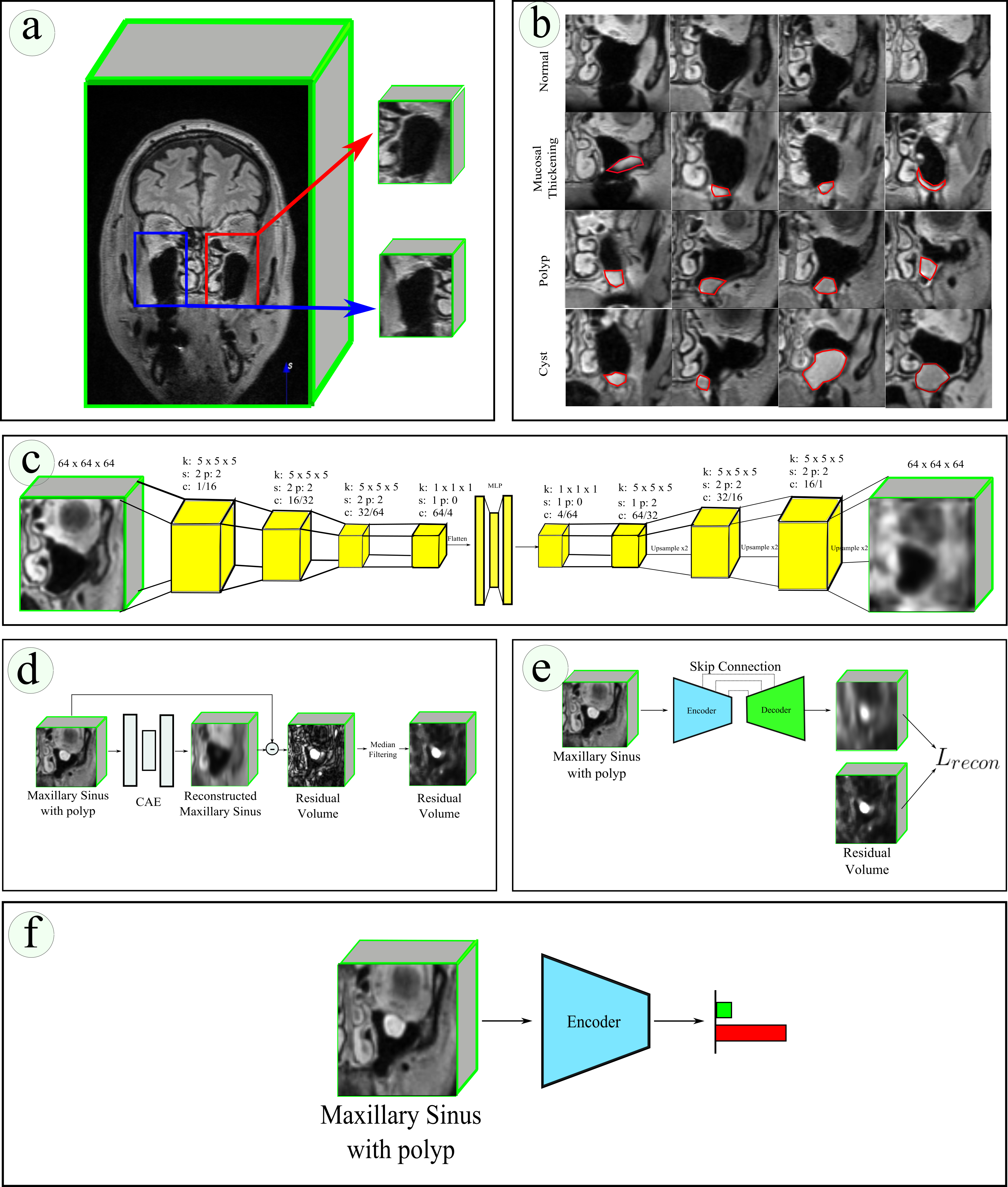} 
  \caption{a) Extraction of MS volumes from cranial MRI b) Exemplarary coronal images of normal MS volume and MS with mucosal thickening, polyp and cyst anomaly c) Our CAE architecture. Here, \textit{k} refers to kernel size, \textit{s} refers to stride, \textit{p} refers to padding, \textit{c} refers to channel where, for example, 1/16 refers to input channel of 1 and output channel of 16. Each stage of the encoder and decoder is formed using 3D convolution followed by batch normlalisation and leaky ReLU. Upsample refers to trilinear upsampling. d) Generation of residual volume required for the self-supervision task using our CAE e) Our self-supervision task where the encoder and decoder is trained to reconstruct the residual volume f) Downstream task where the self-supervision trained encoder is trained to classify between normal and anomalous MS. } 
  \label{fig:method}
\end{figure}

\begin{figure}[htpb!]
  \centering
  \includegraphics[width=\columnwidth]{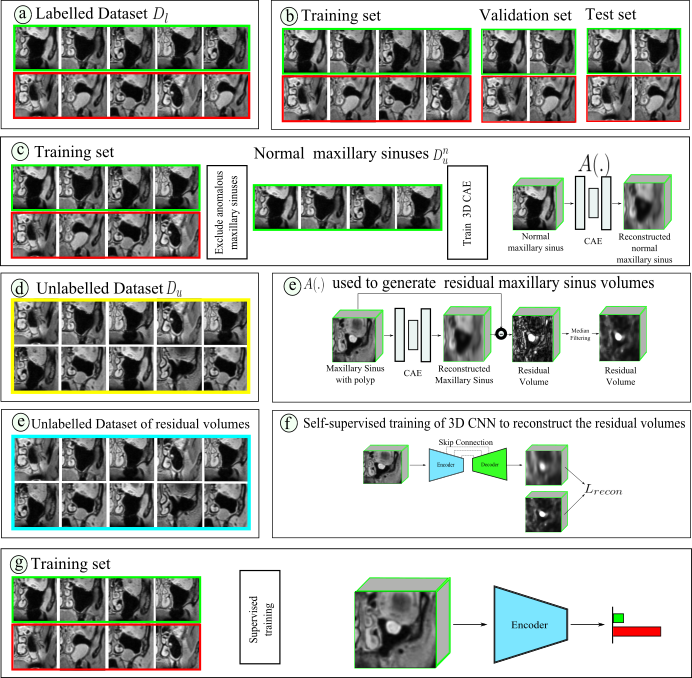} 
  \caption{ Our data processing pipeline comprises several steps: a) The labelled dataset \(D_l\) b) Splitting \(D_l\) into training, validation, and test subsets for downstream classification of normal versus anomalous MS. c) Normal MS samples from the labelled training set form \(D_{l}^{n}\), used to train the 3D CAE \(A(.)\) within the UAD framework. d) Unlabelled dataset \(D_u\) e) This trained 3D CAE \(A(.)\) generates residual volumes from the unlabelled dataset \(D_u\) e)  Unlabelled dataset of residual volumes f) The 3D CNN undergoes self-supervised training to reconstruct these residual volumes. g) The 3D CNN's encoder is initialized with weights from the SSL task, then undergoes supervised training for the final task of classifying normal versus anomalous MS, using the training set created in step a). } 
  \label{fig:dataproc}
\end{figure}

\subsection{Description of Dataset}


As part of the Hamburg City Health Study (HCHS) \cite{Jagodzinski2019-ct}, cranial MRI scans were obtained from individuals aged 45-74 years to evaluate neuroradiological parameters. The scans were acquired using fluid attenuated inversion recovery (FLAIR) sequences in the NIfTI format at the University Medical Center Hamburg-Eppendorf. The MRI scans had a resolution of 173 mm x 319 mm x 319 mm. The labelled dataset consisted of 1067 patients. Among the patients, 489 exhibited no pathologies in their left and right MS, while 578 had at least one MS presenting polyp, cyst or mucosal thickening pathology.  All these anomalies were grouped into the "anomaly" class. Our unlabelled dataset consists of 1559 patient MRIs.  The diagnoses were established by two ENT specialists and one radiologist specialized in ENT.  Figure \ref{fig:method} b shows coronal slices highlighting the diverse set of anomalies that are present in our dataset.

\subsection{Dataset preprocessing}

In our dataset preprocessing, as outlined in previous work \cite{bhattacharya2023multiple,10.1007/978-3-031-16437-8_41}, we first align MRIs with a fixed sample from our dataset. Centroid locations of left and right MS regions were recorded for 20 patients, guiding the extraction of MS volumes from larger cranial MRIs. This step isolates the relevant MS volumes for our task of classifying healthy and anomalous MS. We then used the mean centroid location from these 20 recordings to extract left and right MS volumes from all cranial MRIs in our dataset. The extracted volumes, sized 64 mm x 64 mm x 64 mm, cover the entire MS. Figure \ref{fig:method} a illustrates this extraction process. 

Each cranial MRI yielded one left and one right MS volume. To enhance symmetry, right MS volumes were horizontally flipped to match the left ones. All volumes were normalized to an intensity range of 0 to 1. We employed five-fold cross-validation for evaluation, ensuring diverse labelled datasets (10\%, 20\%, 40\%, 60\%, 80\%) maintain the anomaly-to-normal ratio. The separation of training, validation, and test sets was strictly maintained, with left or right MS volumes from the same patient assigned to only one set. Table \ref{tab:dataset} details our dataset division across these sets.

\begin{table}[h]
\centering
\begin{tabular}{|c|c|c|c|}
\hline
\textbf{Class} & \textbf{Training Set} & \textbf{Validation Set} & \textbf{Test Set} \\
\hline
\# Normal MS & 708 & 176 & 380 \\
\hline
\# Anomalous MS & 487 & 122 & 261 \\
\hline
\end{tabular}
\caption{Statistics of our labelled dataset \(D_l\)}
\label{tab:dataset}
\end{table}

\subsection{Architecture}

Our CAE, depicted in Figure \ref{fig:method} c, uses 3D convolutional operations with a latent bottleneck dimension of 512. The CNN architecture is U-Net inspired, featuring a 3D ResNet18 encoder \(E(.)\) \cite{8578773} with four stages and channel dimensions of 64, 128, 256, and 512. The decoder \(D(.)\) mirrors the encoder, with reverse channel dimensions and trilinear upsampling. Skip connections are used to pass encoder features to the decoder. For Bootstrap your own latent (BYOL), SimSiam, and SimCLR training, only the encoder \(E(.)\) is used, with an MLP attached to project the final layer features to a dimension of 512.

\subsection{Autoencoder training and inferrence on unlabelled dataset}

Consider \(D_{l}\) to be our labelled dataset containing normal and anomalous MS and \(D_{u}\) to be our unlabelled dataset. Further, let \(D_{l}^{n} \subset D_{l}\) be a dataset consisting of only normal MS volumes. Let \(x \in R^{64 \times 64 \times 64}\) be an MS volume in \(D_{l}\). Let the autoencoder be represented as \(A(.)\) such that \(x' = A(x)\) represents the reconstructed MS volume. We train the autoencoder using L1 reconstruction loss which may be written as \(\|x-x'\|\) on \(D_{l}^{n}\). Once trained, we use the autoencoder \(A(.)\) to generate residual volumes on \(D_{u}\). Figure \ref{fig:method} d illustrates our residual volume generation method.

\subsection{Transfer Learning} 

\textcolor{black}{Since transfer learning (TL) is a method to achieve data efficiency, we also trained our models initialised with transfer learning weights. However, since our downstream task involves MRI and is in 3D domain, ImageNet \cite{5206848} weights may not be appropriate. Hence, the model weights we utilized as initial weights were obtained through training on eight diverse public 3D segmentation datasets, covering both MRI and CT modalities. We believe these weights are more suitable than those derived from natural image training and therefore employed them as the basis for our 3D CNN. For further information on the transfer learning model, please see the GitHub repository \footnote{https://github.com/Tencent/MedicalNet}.} 

\subsection{Self-Supervised Training} 
With the residual volumes generated for \(D_{u}\), we train \(E(.)\) and \(D(.)\) to reconstruct the residual volumes again. This, in effect, makes the encoder and decoder learn features relevant for anomaly localisation within the unlabelled dataset  \(D_{u}\). We train \(E(.)\) and \(D(.)\) using \(L_{recon}\) which in our case is binary cross entropy (BCE) loss. Figure \ref{fig:method} e illustrates our self-supervised training task. 
We evaluated our self-supervised learning method against Autoencoder (AE), Denoising Autoencoder (DAE), BYOL, SimSiam, SimCLR \textcolor{black}{and Sparse
masKed modeling with hierarchy (SparK)}. These methods use similar encoders \(E(.)\) and decoders \(D(.)\), with BYOL, SimSiam, and SimCLR employing an additional MLP for feature projection. \textcolor{black}{Pretraining with the SparK framework requires sparse encoder \(E'(.)\) and a special light decoder which contains 3 convolutional blocks and 3 upsampling blocks \cite{tian2023designing}. Patch size 8 \(\times\) 8 \(\times\) 8 and masking ratio of 60\% was used during pretraining}. Detailed description and implementation details of our state-of-the-art (SOTA) SSL methods is provided in the supplementary material section 1-7. \textcolor{black}{More details about the other masking ratios and patch sizes tested for SparK can be found in the supplementary material section 11.}

\subsection{Finetuning}

Having successfully trained the \(E(.)\) and \(D(.)\) using self-supervision, we move onto the finetuning phase. We discard \(D(.)\) and focus on training \(E(.)\) by leveraging samples from the labelled dataset \(D_{l}\). \textcolor{black}{For TL models, we initialise \(E(.)\) with transfer learning weights.} Next, we introduce a MLP as an additional component, responsible for projecting the encoder features from their original dimension of 512 to an intermediate dimension of 256. Subsequently, the MLP maps these features to a final dimension of 2, corresponding to the number of classes. We finetune \(E(.)\) using BCE loss.  

Figure \ref{fig:dataproc} illustrates the data processing pipeline and elucidates how the different components fit into our overall method. 

\subsection{Implementation details} 

Our PyTorch and PyTorch Lightning-based code accommodates a maximum batch size of 256 on NVIDIA A6000 with 48GB VRAM for self-supervised pretraining. We optimize models using LARS \cite{ginsburg2018large} with a learning rate of 0.2 across 500 epochs, incorporating a 20-epoch linear warmup and cosine annealing. For finetuning, AdamW \cite{loshchilov2018decoupled} is employed with a constant rate of 1e-4 for 100 epochs at a batch size of 16. Models yielding the lowest validation loss are preserved for final evaluation with the test set. The CAE was trained on 708 normal MS volume samples without augmentation. For self-supervised methods and MS anomaly classification, we applied data augmentations such as random affine transformations, flipping, and Gaussian noise. The DAE specifically used Gaussian noise with a mean of 0 and standard deviation of 0.6 at 100\% probability, while other augmentations were applied 50\% of the time. Supplementary material offers comprehensive descriptions and visualizations of SOTA SSL methods.

\section{Results}

\textbf{Comparison to state of the art}

\begin{figure}[!htb]
    \centering
    \includegraphics[width=\columnwidth]{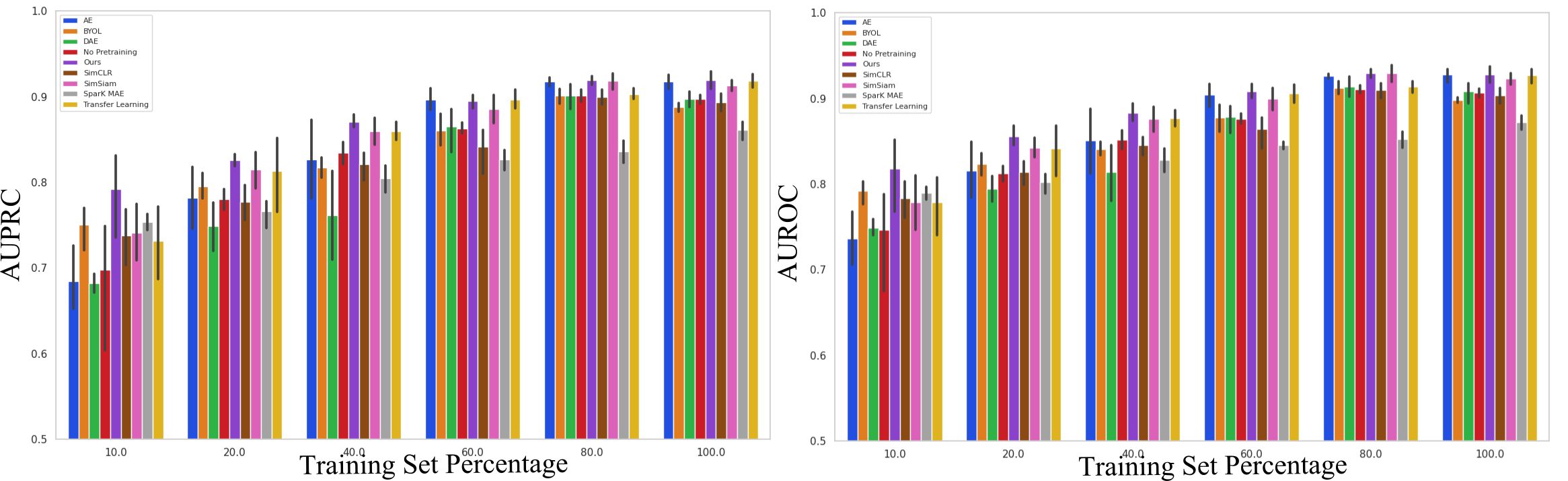}
    \caption{ (LEFT) AUPRC trend vs training set percentage (RIGHT) AUROC trend vs training set percentage  }
    \label{fig:result_A}
\end{figure}

Results in Table \ref{tab:my-table} show our method outperforming others in AUROC, AUPRC, and F1 scores across different labelled dataset scenarios (10\%, 20\%, 100\% of \(D_{l}\)). Our method demonstrated notable improvements in AUROC (3.34\% and 4.93\% over SimSiam) and AUPRC (5.33\% over BYOL and 5.12\% over AE) for 10\% and 20\% dataset scenarios, respectively. \textcolor{black}{SparK trained models perform generally poorer compared to the other SSL and TL methods with the performance gap between SparK MAE and our method widening with increased training set percentage.  Our method had AUPRC 8.21\% higher than the TL method when finetuned on a 10\% training set.} Pretraining models using our method significantly boosted AUPRC by 14.49\% and AUROC by 9.45\% compared to no pretraining when trained on a 10\% training dataset. At 100\% dataset finetuning, our method achieved the highest scores, with AE and SimSiam showing similar performance. Compared to no pretraining, our method improved AUPRC by 3.33\%. Figure \ref{fig:result_A} illustrates AUPRC and AUROC trends with increasing training set percentages, respectively. Our method excels in settings with 40\% or less training data but aligns with SOTA performance beyond that. 

\begin{table}[htbp]
\centering
\caption{
The table displays the mean and 95\% confidence intervals of metrics evaluating model performance in the downstream classification task. These models, trained with varying portions of \(D_l\), were initialized using different SSL methods before supervised training.}

\label{tab:my-table}
\resizebox{\textwidth}{!}{%
\begin{tabular}{|c|c|c|c|c|}
\hline
\textbf{Method} & \textbf{Training Set Percentage \(D_l\)} & \textbf{AUROC} & \textbf{AUPRC} & \textbf{F1} \\ \hline
No pretraining & 10\% & 0.74 (0.64-0.84) & 0.69 (0.56-0.82)  & 0.64 (0.59-0.69)\\ 
Transfer Learning & 10\% & 0.77 (0.72-0.82) & 0.73 (0.66-0.79)  & 0.63 (0.57-0.69)\\ 
AE & 10\% & 0.73 (0.68-0.79) & 0.68 (0.62-0.74) &  0.55 (0.43-0.67)\\ 
DAE & 10\% & 0.74 (0.73-0.76) & 0.68 (0.66-0.69) &  0.62 (0.60-0.64)\\ 
BYOL & 10\% & 0.79 (0.76-0.81) & 0.75 (0.70-0.79) &  0.63 (0.59-0.69)\\ 
SimSiam & 10\% & 0.77 (0.72-0.83) & 0.74 (0.68-0.79) & 0.62 (0.53-0.72)  \\ 
SimCLR & 10\% & 0.78 (0.74-0.81) & 0.73 (0.68-0.78) & 0.63 (0.59-0.68)\\ 
SparK MAE & 10\% & 0.78 (0.77-0.80) & 0.75 (0.73-0.76) & 0.65 (0.63-0.67)\\ 
Ours & 10\% & \textbf{0.81 (0.74-0.88)} & \textbf{0.79 (0.71-0.87)} & \textbf{0.67 (0.58-0.77)} \\ 
\hline
No pretraining & 20\% & 0.81 (0.79-0.82) & 0.78 (0.76-0.79) & 0.67 (0.65-0.69)\\ 
Transfer Learning & 20\% & 0.84 (0.79-0.88) & 0.81 (0.74-0.88)  & 0.68 (0.62-0.75) \\
AE & 20\% & 0.81 (0.76-0.86) & 0.78 (0.72-0.83) &  0.67 (0.60-0.74)\\
DAE & 20\% & 0.79 (0.77-0.81) & 0.74 (0.70-0.79) & 0.67 (0.64-0.70)\\
BYOL & 20\% & 0.82 (0.80-0.84) & 0.79 (0.77-0.82) & 0.70 (0.68-0.71)\\
SimSiam & 20\% & 0.84 (0.82-0.86) & 0.81 (0.78-0.84) & 0.70 (0.67-0.74)\\ 
SimCLR & 20\% & 0.81 (0.79-0.83) & 0.77 (0.74-0.81) & 0.68 (0.67-0.69)\\ 
SparK MAE & 20\% & 0.80 (0.78-0.82) & 0.76 (0.73-0.79) & 0.67 (0.65-0.68)\\ 
Ours & 20\% & \textbf{0.85 (0.83-0.87)} & \textbf{0.82 (0.81-0.83)}  & \textbf{0.72 (0.70-0.75)}\\ 
\hline
No pretraining & 100\% & 0.90 (0.89-0.91) & 0.89 (0.88-0.90)  & 0.80 (0.78-0.82)\\ 
Transfer Learning & 100\% & 0.92 (0.91-0.93) & 0.91 (0.90-0.93)  & 0.82 (0.80-0.83) \\ 
AE & 100\% & 0.92 (0.91-0.93) & 0.91 (0.90-0.93)  & 0.82 (0.80-0.84)\\ 
DAE & 100\% & 0.90 (0.88-0.92) & 0.89 (0.88-0.91) & 0.79 (0.77-0.82)\\ 
BYOL & 100\% & 0.89 (0.89-0.90) & 0.88 (0.87-0.89) & 0.78 (0.76-0.81)\\ 
SimSiam & 100\% & 0.92 (0.91-0.93) & 0.91 (0.90-0.92)  & 0.81 (0.79-0.83)\\ 
SimCLR & 100\% & 0.90 (0.88-0.91) & 0.89 (0.87-0.91) & 0.79 (0.77-0.80)\\ 
SparK MAE & 100\% & 0.87 (0.85-0.88) & 0.86 (0.84-0.87) & 0.75 (0.73-0.76)\\ 
Ours & 100\% & \textbf{0.93 (0.91-0.94)} & \textbf{0.92 (0.90-0.93)}  & \textbf{0.83 (0.80-0.86)}\\ \hline

\end{tabular}
}
\end{table}

\textbf{Effect of varying the CAE training set }

The effectiveness of our self-supervised task is contingent on the CAE's proficiency in reconstructing healthy MS volumes. Inaccurate reconstructions yield unreliable residuals, affecting self-supervision. To assess the impact of training set size, the CAE was trained with different proportions (20\%, 40\%, 60\%, 80\%, 100\%) of the healthy MS dataset \(D_{l}^{n}\). After training, the CAE processed dataset \(D_{u}\) to produce residual volumes, which were refined using a median filter with a kernel size of 5. Subsequent supervised training utilized 10\% of our labelled dataset \(D_{l}\). Table \ref{tab:autoencfn} presents improvements in the downstream task metrics correlating with increased healthy MS training set sizes, suggesting that larger \textit{normal} dataset \(D_{l}^{n}\) enhance normal MS representation learning and improve anomaly localization.

\begin{table}[htbp!]
\centering
\caption{The table shows the mean and 95\% confidence intervals of metrics for evaluating model performance in downstream classification. The CAE was trained on varying proportions of the normal MS volumes dataset (\(D_{l}^{n}\)), then used to generate residual volumes from the unlabelled dataset (\(D_u\)). Each model was initialized using our proposed SSL method.}
\label{tab:autoencfn}
\resizebox{\textwidth}{!}{%
\begin{tabular}{|c|c|c|c|}
\hline
\textbf{Training Set Percentage \(D_{l}^{n}\)}  & \textbf{AUROC} & \textbf{AUPRC} & \textbf{F1}  \\ \hline
20\% & 0.76 (0.70-0.81) & 0.72 (0.67-0.77) & 0.60 (0.50-0.69)   \\  
40\% & 0.77 (0.73-0.80) & 0.72 (0.66-0.78) & 0.63 (0.57-0.68)  \\ 
60\% & 0.78 (0.75-0.80) & 0.74 (0.71-0.77)& 0.65 (0.62-0.68)  \\ 
80\% & 0.80 (0.76-0.84) & 0.76 (0.72-0.81)  & 0.67 (0.63-0.72)  \\ 
100\% & \textbf{0.81 (0.74-0.88)} & \textbf{0.79 (0.71-0.87)}  & \textbf{0.67 (0.58-0.77)} \\ 

\hline

\end{tabular}
}
\end{table}
\section{Discussion}

Tailoring SSL tasks to specific downstream tasks offers distinct advantages \cite{ozbulak2023know}. Current SOTA SSL methods \cite{NEURIPS2020_f3ada80d,9578004,Huang2023}, primarily developed for 2D image classification on datasets like ImageNet, do not address the unique challenges of 3D MRI modalities and the specifics of paranasal anomalies. Our SSL task is specifically tailored to address the challenges associated with 3D environments, MRI modality, and the classification of paranasal anomalies. 

 We conjecture that segmentation of anomalies as a SSL task, requiring knowledge of anomaly locations, enhances the learning of class-discriminative features for distinguishing normal and anomalous MS. Our SSL task is a segmentation task therefore, it requires segmentation masks highlighting anomalies. To avoid the high costs of annotation, we use a CAE trained in the UAD framework for generating approximate annotations, effective in localizing paranasal anomalies \cite{https://doi.org/10.48550/arxiv.2211.01371}. This CAE training utilizes labelled \textit{normal} datasets, typically accessible in supervised settings. Unlike generic SOTA SSL methods, which do not prioritize anomaly localization, our approach demonstrates improved AUROC and AUPRC (as shown in Table \ref{tab:my-table}), suggesting that effective anomaly localization can enhance classification performance, even with limited labelled data. Methods like BYOL and SimSiam, which aim to maximize agreement between augmented views, are less effective for paranasal anomaly classification. SimCLR’s performance shortfall is likely due to smaller batch sizes, a necessity given the impracticality of large batches in 3D settings, despite SimCLR's recommendation of 4096 \cite{10.5555/3524938.3525087}. Our method is more suited for such constrained computational resources. AE and DAE, focusing on compression-decompression and denoising, do not guarantee discriminative feature learning for downstream classification \cite{10.1145/1273496.1273592}, and were found less effective in our context. When the entire training set is used, our method, AE, and SimSiam yield comparable results, with ours marginally outperforming. \textcolor{black}{We also explored MAE-style pretraining using SparK. However, the results suggest that fine-tuning performance is notably weaker, particularly when fine-tuning with a training set percentage 40\% and above. These findings imply that generating masked regions contributes to representation learning; however, the acquired representations do not appear to enhance downstream classification. It is noteworthy that the SparK framework was initially developed and evaluated for 2D natural images. Although we adapted the framework for 3D applications, our findings underscore the necessity for further methodological advancements to effectively support tasks in the 3D domain. Further, TL models exhibit comparable performance to SSL methods when fine-tuning on training sets exceeding 20\%. This suggests that transfer learning methods remain viable for paranasal anomaly classification given an ample supply of labeled samples. However, in the scenario of an extremely limited labeled dataset, such as 10\%, our method outperforms TL, indicating that the representations acquired by our approach are especially advantageous in low-data environments.} Overall, compared to approaches without pretraining, our tailored SSL task consistently shows superior downstream classification performance, underlining its efficacy.


Our analysis regarding the impact of the CAE training set size shown in Table \ref{tab:autoencfn} has demonstrated that the inclusion of a substantial cohort of normal MS volumes yields notable benefits for both the self-supervision task and the subsequent downstream task suggesting that better anomaly localisation by the CAE and thereby better representation learning by the CNN in the self-supervision task. We also analysed the influence of the loss function and post-processing used in the self-supervision task which can be found in the supplementary material section 8 and 9. 

Our study has limitations that require further investigation. It is based on a single-center, MRI-only study, so multi-center studies with varied imaging modalities are needed for generalizability. Our methods rely on a cohort of healthy MS volumes, unlike other self-supervised tasks. We focused on convolutional autoencoders, not exploring models like variational autoencoders 
Generative Adversarial Networks, 
 or transformer-based architectures 
and diffusion models 
, which might offer better anomaly localization. We compared L1, L2, and BCE loss functions but not others like the Structural Similarity Index 
or perceptual loss. 
Future research should examine these aspects and apply this self-supervision approach to other domains, like brain anomaly detection.

\section{Conclusion}

We developed a novel self-supervision task that focuses on anomaly localization to better classify paranasal anomalies in the maxillary sinus, addressing the lack of methods that effectively use unlabelled datasets to learn discriminative features for this purpose. Our approach uses an autoencoder trained on healthy MS volumes to generate residual volumes from an unlabelled dataset. These residuals serve as coarse segmentation masks for localizing anomalies. By training a CNN to reconstruct these volumes, it implicitly learns anomaly localization, thereby developing transferable features for the downstream classification task. Our method outperforms existing self-supervision techniques, proving its effectiveness in this specific domain.

\bmhead{Acknowledgements}

This work has not been submitted for publication anywhere else. This work is funded partially by the i3 initiative of the Hamburg University of Technology. The authors also acknowledge the partial funding by the Free and Hanseatic City of Hamburg (Interdisciplinary Graduate School) from University Medical Center Hamburg-Eppendorf. This work was partially funded by Grant Number KK5208101KS0 (Zentrales Innovationsprogramm Mittelstand, Arbeitsgemeinschaft industrieller Forschungsvereinigungen).

\backmatter

\bmhead{Ethical approval declarations} The study protocol received approval from the local ethics committee (Landesärztekammer Hamburg, PV5131) and was approved by the Data Protection Commissioners for the University Medical Center of the University Hamburg-Eppendorf and the Free and Hanseatic City of Hamburg. It is registered on ClinicalTrial.gov (NCT03934957) and adheres to Good Clinical Practice, Good Epidemiological Practice, and ethical principles outlined in the Declaration of Helsinki.

\bmhead{Conflicts of Interest}

The authors declare no conflict of interest.

\bibliography{sn-bibliography}


\end{document}